\documentclass[3p,times,procedia]{elsarticle}
\usepackage{nupha_ecrc}
\usepackage[belowskip=-10pt,aboveskip=5pt]{caption}

\volume{00}

\firstpage{1}

\journalname{Nuclear Physics A}

\runauth{A. Beraudo et al.}

\jid{nupha}

\jnltitlelogo{Nuclear Physics A}

\usepackage{amssymb}
\newcommand{\beq}{\begin{equation}}
\newcommand{\eeq}{\end{equation}}

\usepackage[figuresright]{rotating}

\begin{document}

\begin{frontmatter}



\dochead{XXVIIth International Conference on Ultrarelativistic Nucleus-Nucleus Collisions\\ (Quark Matter 2018)}

\title{Heavy-flavor flow-harmonics in high-energy nuclear collisions: time-development and eccentricity fluctuations}


\author{A. Beraudo, A. De Pace, M. Monteno, M. Nardi and F. Prino}

\address{INFN - Sezione di Torino, Via P. Giuria 1, I-10125 Torino}

\begin{abstract}
We study the development of heavy-flavor flow harmonics in high-energy nuclear collisions. The elliptic and triangular flow of heavy-flavor hadrons, arising from the finite impact parameter of the two nuclei and from event-by-event fluctuations of the initial geometry, is analyzed in detail, considering the contribution from particles decoupling from the fireball at various times. We also study the dependence of the flow harmonics on the event-shape fluctuations, considering events belonging to the same centrality class but characterized by very different eccentricities (or vice-versa).   
\end{abstract}

\begin{keyword}
Quark-Gluon Plasma \sep Heavy Quarks \sep Flow harmonics \sep Event-shape engineering

\end{keyword}

\end{frontmatter}


\section{Introduction}
Heavy quarks in relativistic heavy-ion experiments are initially produced off-equilibrium in hard pQCD processes and, before hadronizing and being detected through their decay products, they cross the deconfined fireball arising from the collision of the two nuclei. Hence, their final distributions allow one to extract information on the properties of the hot QCD plasma crossed during their propagation, in particular on its transport coefficients. The first observables addressed in Heavy-Flavor (HF) studies were the nuclear modification factor $R_{\rm AA}$ and the elliptic-flow coefficient $v_2$ in non-central collisions, arising mainly from the finite impact parameter. More refined experimental analysis and theoretical studies, accounting for event-by-event fluctuations in the initial energy deposition and eccentricity, have the potential to provide a richer information both on the initial state of the collision and on the interaction of the heavy quarks with the medium. This is the subject of the present contribution focused on recent developments achieved with the POWLANG transport setup~\cite{Beraudo:2017gxw}.
\section{Development of elliptic and triangular flow}
\begin{figure}[!ht]
\begin{center}
  \includegraphics[clip,width=0.4\textwidth]{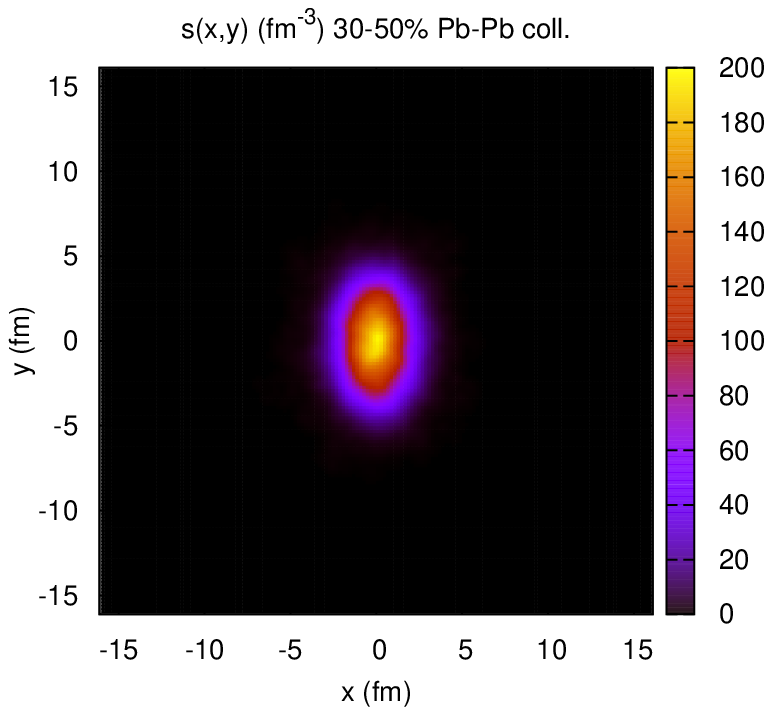}
  \includegraphics[clip,width=0.4\textwidth]{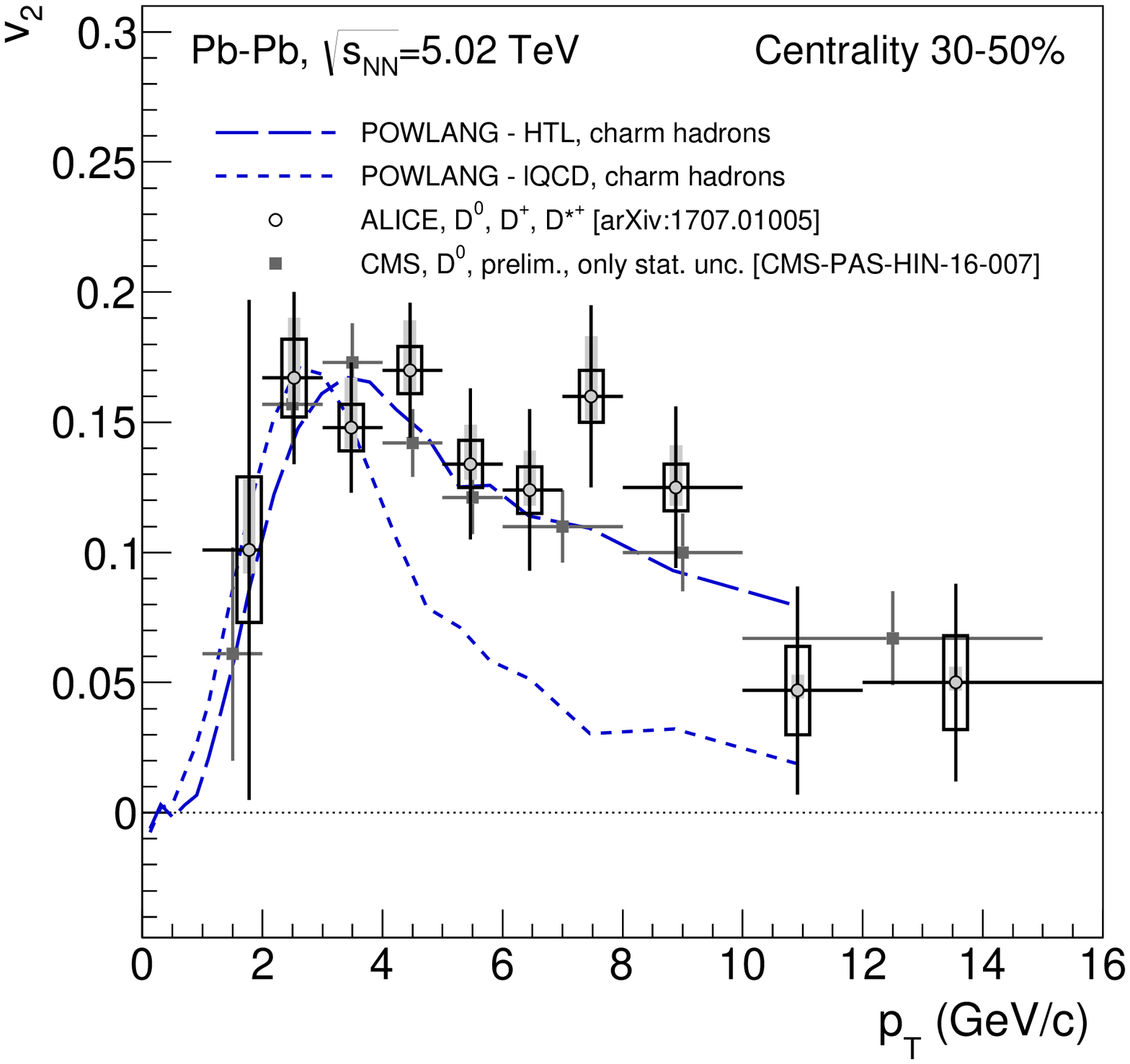}
  \caption{Initial condition for non-central Pb+Pb collisions with an elliptic deformation and the resulting $v_2$ coefficient for $D$-mesons compared to ALICE~\cite{Acharya:2017qps} and CMS data~\cite{Sirunyan:2017plt}.}\label{fig:e2v2}
\end{center}
\end{figure}

\begin{figure}[!ht]
\begin{center}
  \includegraphics[clip,width=0.4\textwidth]{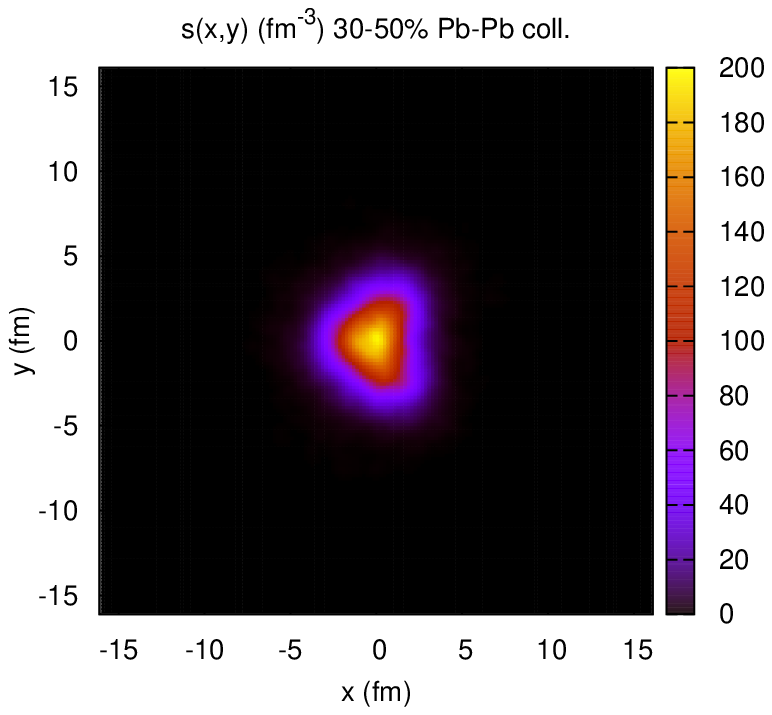}
  \includegraphics[clip,width=0.4\textwidth]{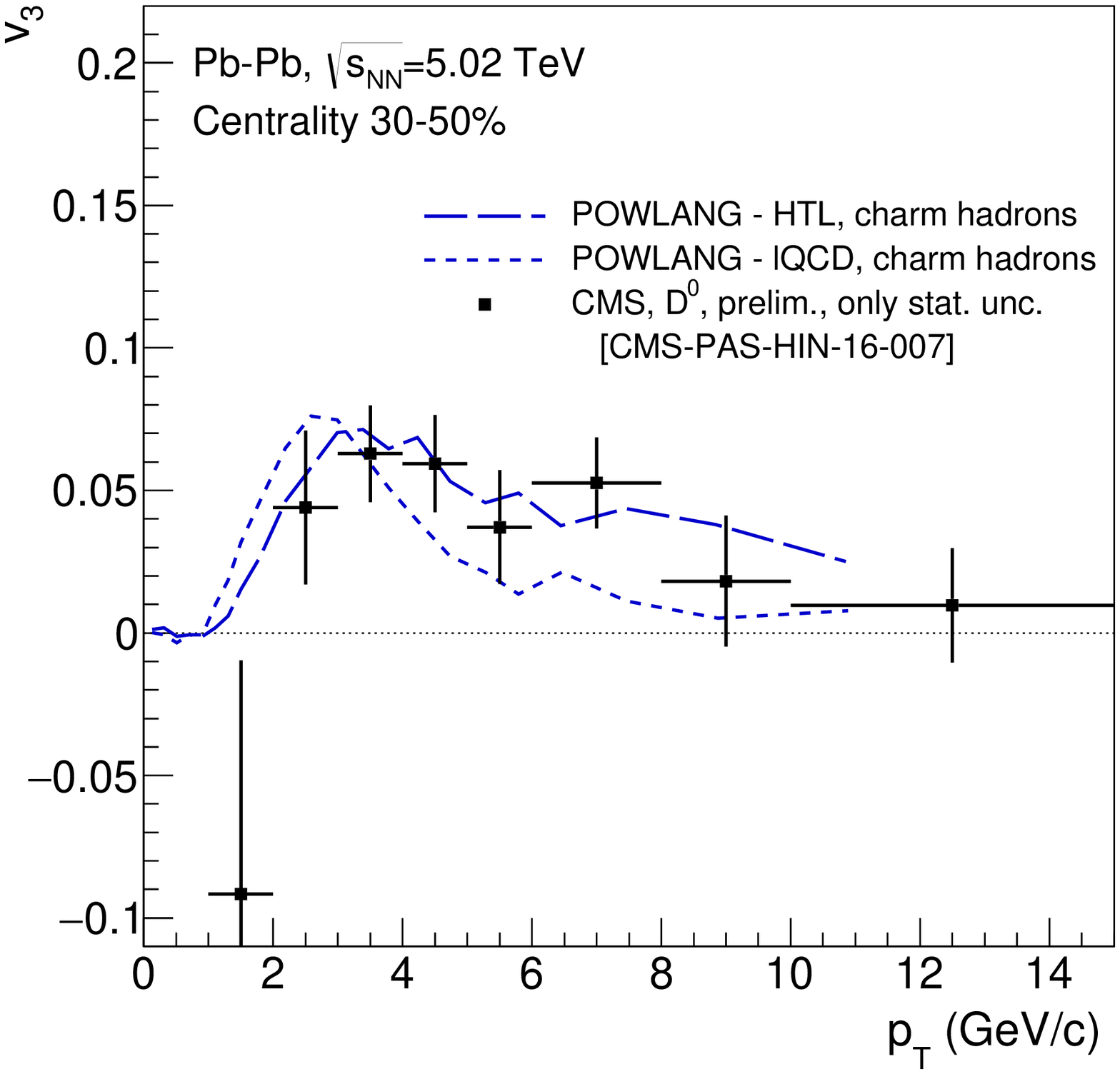}
\caption{Initial condition for non-central Pb+Pb collisions with a triangular deformation and the resulting $v_3$ coefficient for $D$-mesons compared to CMS data~\cite{Sirunyan:2017plt}.}\label{fig:e3v3}
\end{center}
\end{figure}

Initial-state event-by-event fluctuations leave their fingerprints in the final azimuthal distributions of hadrons: non-zero values of the Fourier coefficients $v_2$ in central collisions and $v_3$ in all centrality classes -- otherwise vanishing for smooth event-averaged initial conditions -- are observed. Here we display the results obtained extending such an analysis to HF particles~\cite{Beraudo:2017gxw}. We proceed as follows. We generate several thousands of Glauber-MC initial conditions, dividing them in centrality classes according to the number of nucleon-nucleon collisions $N_{\rm coll}$. In a given event each collision is assumed to deposit some entropy in the transverse plane, with a Gaussian smearing, and the resulting anisotropy can be quantified by the coefficients
\beq
\epsilon_m e^{im\Psi_m}\equiv-\left\{r_\perp^2e^{im\phi}\right\}/{\{r_\perp^2\}}.
\label{eq:eccpsi}
\eeq
In the above curly brackets refer to an average over the transverse plane weighted by the local entropy density.
Exploiting the linear response of the lowest-order flow coefficients to the initial geometric eccentricity $v_{2/3}\!\sim\! \epsilon_{2/3}$, for each centrality class we build an average initial condition summing all the events, each one rotated so to have the reference angle $\Psi_{2/3}$ aligned along the $x$-axis. The hydrodynamic evolution of the fireball is then calculated through the ECHO-QGP code~\cite{DelZanna:2013eua} and the propagation of the heavy quarks throughout the medium is simulated via the Langevin equation implemented in the POWLANG setup, which includes also a routine modeling in-medium hadronization. Results for the $D$-meson $v_2$ and $v_3$ in non-central Pb+Pb collisions at the LHC~\cite{Beraudo:2017gxw} are displayed in Figs.~\ref{fig:e2v2} and \ref{fig:e3v3} and compared to ALICE~\cite{Acharya:2017qps} and CMS~\cite{Sirunyan:2017plt} data. For a similar theoretical study of the HF triangular flow see~\cite{Nahrgang:2014vza}.   

\begin{figure}[!ht]
\begin{center}
  \includegraphics[clip,height=5cm]{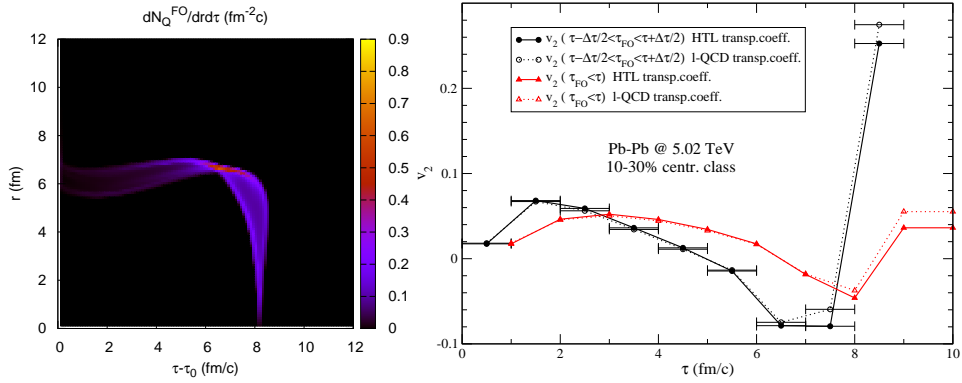}
  \includegraphics[clip,height=5cm]{v2-vs-FO_comp_10-30.eps}  
\caption{The distribution of charm quarks decoupling from the various fluid cells (left panel) and the time-development of their azimuthal elliptic anisotropy.}\label{fig:v2vstau}
\end{center}
\end{figure}

Besides getting values for the HF flow coefficients in agreement with the experiment, it is of interest to understand their origin, i.e. whether they reflect a certain degree of thermalization of the heavy quarks with the surrounding plasma of light partons or other effects, like an anisotropic escape probability of the heavy quarks produced near the edge of the fireball or the different medium-length crossed at the various azimuthal angles. The results of such an analysis are displayed in Fig.~\ref{fig:v2vstau}, referring to the case of the charm $v_2$. The final signal comes from the interplay of different (opposite-sign) effects: at the very beginning it is dominated by the anisotropic escape probability and only later the interaction with the medium plays a role. 

\begin{figure}[!ht]
\begin{center}
  \includegraphics[clip,height=6cm]{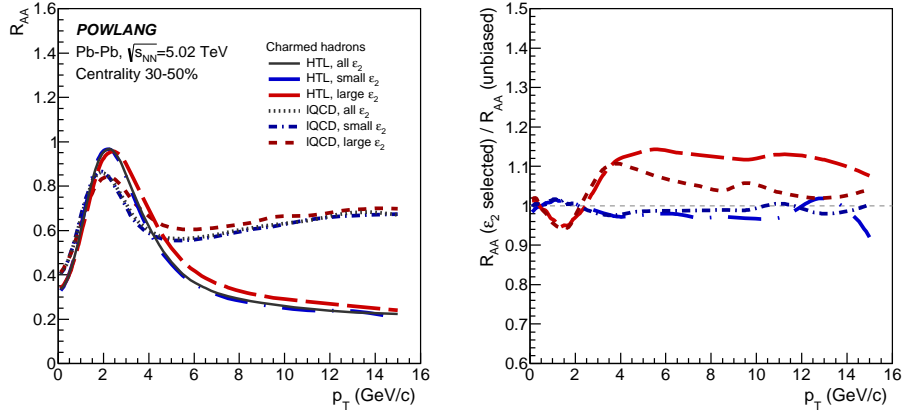}  
\caption{The nuclear modification factor of charmed hadrons in 30-50\% Pb+Pb collisions at $\sqrt{s_{\rm NN}}\!=\!5.02$ TeV for different eccentricity selections. For both choices of transport coefficients the results display only a mild sensitivity to the initial eccentricity.}\label{fig:ESE_RAA}
\end{center}
\end{figure}
\begin{figure}[!ht]
\begin{center}
  \includegraphics[clip,height=6cm]{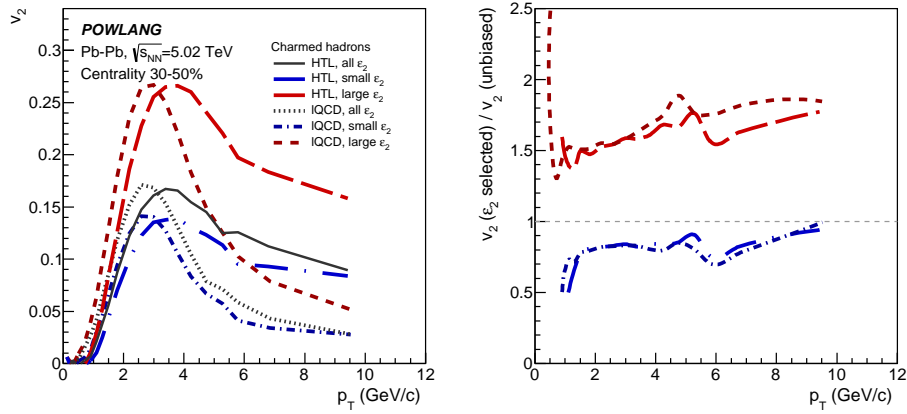}  
\caption{The charmed hadron elliptic flow in 30-50\% Pb+Pb collisions at $\sqrt{s_{\rm NN}}\!=\!5.02$ TeV for different eccentricity selections. The response to the initial eccentricity fluctuations is strong and of the same amount for both sets of transport coefficients.}\label{fig:ESE_v2}
\end{center}
\end{figure}
\section{Event-shape engineering}
Events belonging to the same centrality class, experimentally identified by some estimator like the number of produced charged hadrons, can be characterized by very different eccentricities and hence values of the azimuthal flow harmonics. It is then of interest to divide the events of a given centrality class in sub-samples corresponding to higher/lower azimuthal anisotropy and to study how the angular (and momentum) distributions of produced particles behave as a function of $p_T$. First studies of such an item mainly concern light hadrons, finding results supporting the picture that the final $v_2$, to a large extent, simply reflect fluctuations in the initial geometry~\cite{Adam:2015eta}.

It is of interest to extend this kind of analysis to HF particles, to understand to what extent their various flow harmonics are simply sensitive to the initial geometric deformation of the fireball or also to the density of the medium, performing then simulations in various centrality classes. In fact, since heavy quarks are initially produced off-equilibrium and -- due to their large mass -- are characterized by longer relaxation times than light partons, interacting with a denser or more dilute medium should affect their response to the initial geometry and to the resulting flow of the fireball.

In such a first study, which is currently work in progress, we proceed as follows. In each centrality class, corresponding to a given range of values of $N_{\rm coll}$, we isolate the subsets of 0-20\% most eccentric and 0-60\% least eccentric events, both in the case of an elliptic and triangular deformation, quantifying the eccentricity of the initial condition through the coefficient, following from Eq.~(\ref{eq:eccpsi}),
\beq
\epsilon_{m}={\sqrt{\{r_\perp^2\cos(m\phi)\}^2+\{r_\perp^2\sin(m\phi)\}^2}}/{\{r_\perp^2\}}.
\eeq
So far, we considered the 0-10\%, 10-30\% and 30-50\% most central events. In the case of the 30-50\% centrality class, for which here we provide a few representative results, with our Glauber-MC simulation of the initial state (for the study of the elliptic flow) this amounts to select the events with $\epsilon_2\!\ge\! 0.58$ and $\epsilon_2\!\le\! 0.50$ for the two subsamples. Results referring to this case are displayed in Figs.~\ref{fig:ESE_RAA} and~\ref{fig:ESE_v2}. We notice that the nuclear modification factor, independently from the choice of the transport coefficients, is only midly sensitive to the initial deformation, as shown in Fig.~\ref{fig:ESE_RAA}. On the contrary, subsamples of events with a different initial eccentricity lead to a very different $D$-mesons $v_2$, as displayed in Fig.~\ref{fig:ESE_v2}. A more detailed analysis will be presented in a forthcoming publication.







\end{document}